# A New Approach for Semantic Web Matching

Kamran Zamanifar[1], Golsa Heidary[2],
Naser Nematbakhsh[1], and Farhad Mardukhi[1]

[1] Dept. of Computer Science, University of Isfahan, Isfahan, Iran
`{zamanifar,nemat,mardukhi}@eng.ui.ac.ir`
[2] Young Researchers Club, Computer Engineering Department,
Islamic Azad University, Najafabad Branch, Iran
`golsa.heidary@gmail.com`

**Abstract.** In this work we propose a new approach for semantic web matching to improve the performance of Web Service replacement. Because in automatic systems we should ensure the self-healing, self-configuration, self-optimization and self-management, all services should be always available and if one of them crashes, it should be replaced with the most similar one. Candidate services are advertised in Universal Description, Discovery and Integration (UDDI) all in Web Ontology Language (OWL). By the help of bipartite graph, we did the matching between the crashed service and a Candidate one. Then we chose the best service, which had the maximum rate of matching. In fact we compare two services` functionalities and capabilities to see how much they match. We found that the best way for matching two web services, is comparing the functionalities of them.

## 1   Introduction

Semantic web is a well defined form of the web in which computer agents are able to use information on the web in the same way as human beings do [7]. In other words, semantic of information is well defined in the semantic web to make automatic knowledge extraction possible. However, semantic web suffers from distributed and heterogeneous information.

We use OWL[1] to drive the semantic and syntactic aspects of a service. In systems which availability is one of the most important factors of quality, when a service crashes, it should be replaced by the most similar service. So a repository of web services by the name of UDDI[2] should be available for choosing a service among the candidate services.

In general, to find a similar service we can measure different features and capabilities of two services. The main features are:

---

[1] Web Ontology Language.
[2] Universal Description, Discovery and Integration.

- Comparison of two services` functionality(capability)
- Comparison of quality of services (QOS)
- Comparison of two services` policies

Capability of a service refers to its functions, but policy focuses on non-functional aspects like availability, response time, reusability… which are considered at design time. One Service may have the same policy or quality as crashed service, but will not be a good choice for substitution. But if the capabilities of two services are similar, with high probability we can say that the two services can be substitute with each other. That's why we compare the functionality of crashed web service with other candidate services.

A web service provider can advertise its web through UDDI. A provider must do its work in two steps: first, he should publish his services which mean that the service should be advertised in a registry. Second, a user`s required service must be available in UDDI, by comparing the service`s capability with advertised one.

Our proposed Semantic Web matching has been done in two phases. In the first phase we compute the matching rate of two web services using by making two individual bipartite graphs. One for input matching and another for output matching. We select the most similar web service in the second phase we choose the service which has the most similarity with the crashed service. The architecture of our work is shown in "fig. 1".

In the remainder of this paper, first in Section 2, we talk about related works, in brief. Then in section 3 we define the main idea of matching semantic Web services, which is computing the matching rate by the help of bipartite graph, and after that we choose the best one .in last part of this section we analyze the time complexity of our algorithm. For better explanation of our algorithm, section 4 describes a simple scenario of matching web services. The discussion about power and weakness of our work in comparison with other works, have become in Section 5. Finally in section 6 the conclusion of our work is given.

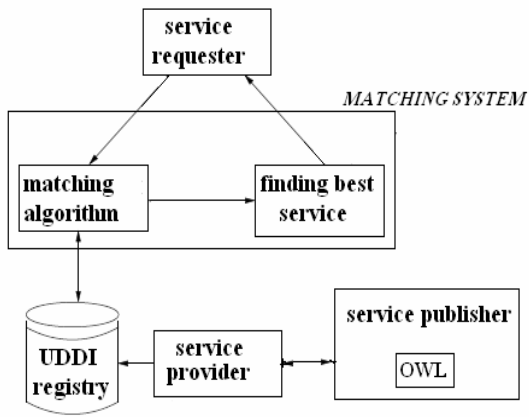

**Fig .1.** Shows the architecture of our matching model

## 2   Related Work

Web services are software building blocks that are in hand through the internet in a standard way using SOAP messaging. The most important properties of web services are reusability and the ability of composition. They allow clients to access data from remote providers without extracting it from HTML web pages or using proprietary protocols. Web services are used in several and different areas such as address validation, trading applications, product search and ordering, automatic systems, and so on. Clients can select from over 100 services listed on providers. Choosing the best service among them which satisfies our need, is as important as difficult. The first choice that comes into mind is Web Service Description Language (WSDL). In theory, semantic information in WSDL files was supposed to solve this problem, because WSDL is a way to know what a service does and how. But in practice, is not enough, because currently WSDL files don't have enough semantic information to decide substitutability or ability of compose. There is a need for automatic techniques to obtain more semantic information.

Many works have addressed ontology matching in the context of ontology design [5],[6],[7]. These works do not deal with explicit notions of similarity. However, many of them [3] have powerful features that allow for smart user interaction, or expressive rule languages for specifying mappings.

In [10] they use ws-policy for matching web services. We know that comparing functionalities will give a better result for semantic web matching.

In [9] semantic Web Service Description Ontology which is called SWSD and QOS, are used to do matching, which are not as good as functionality matching. In this work also, comparing service functionalities are considered in brief.

The similarity measured in [12], is based on statistics, but our work is precise, not on probability.

References [2],[11] have discussed about functional matching of semantic web services, but because the main idea of these works is about composition of web services, matching had not discussed very well.

## 3   Finding Most Similar Service

We have a repository of services that are all described in the same ontology language, OWL. So we can easily compare the capability and functionality of them. Each service does a special work by the means of some functions. These functions have inputs and outputs. Therefore, in order to compare two web services` functionalities, we must compare inputs and outputs. Now we describe the two phases of matching.

### 3.1   Computing the Matching Rate

The first phase is computing the similarity rate of two services. If the crashed service is C and the advertised service is ADV the inputs of each one is shown by $C_{in}$ and $ADV_{in}$ and the outputs are shown by $C_{out}$ and $ADV_{out}$. So we compare $C_{in}$ with $ADV_{in}$ and $C_{out}$ with $ADV_{out}$. Four results will be achieved that the algorithm is as follows:

```
Case (Cout, ADVout):
      If Cout= ADVout then
            return exact
      If Cout, subclass of ADVout then
            return exact
      If ADVout subsumes Cout then
            return plugin
      If Cout subsumes ADVout then
            return subsumes
      Otherwise
            return fail
```

Between all services in the repository and crashed service, this comparison should be done. An Equivalent algorithm also used for inputs.

### 3.1.1  Using Bipartite Graph

We do the matching by the help of bipartite graph. A Bipartite Graph is a graph $G = (V,E)$ in which the vertex set can be partitioned into two disjoint sets, $V = V_0 + V_1$, such that every edge e in E has one vertex in $V_0$ and another in $V_1$. The matching is complete if and only if, all vertices in $V_0$ are matched. It means that all vertices in $V_0$, as well as $V_1$, should have an edge.

Let $C_{out}$ and $ADV_{out}$ be the set of output concepts in C and ADV respectively. These constitute the two vertex sets of our bipartite graph. Construct graph $G=(V_0 + V_1, E)$, where, $V_0 = C_{out}$ and $V_1 = ADV_{out}$. Consider two concepts a in $V_0$ and b in $V_1$ It means that a is one of the output parameters of C and b is one of the output parameters of ADV. Let R be the result of CASE (in our algorithm, which can be Exact=E, Plugin=P, Subsume=S, Fail=F) between concepts a and b. It is obvious that E > P > S > F. We define an edge (a, b) in the graph and label this edge as R. Therefore if matching is complete (all vertices have at least one edge), now we compute the whole matching rate for these two services. In "fig. 2" we have an example of a bipartite graph which has the complete matching.

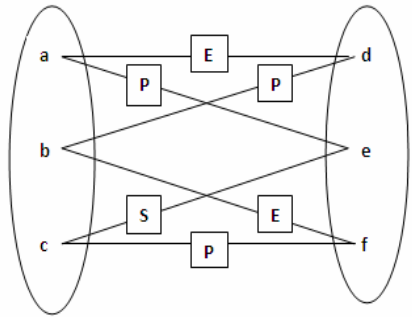

**Fig. 2.** Shows an example of bipartite graph of output concepts

Now we should choose the best sub graph. So for each vertex in $V_0$ we should choose the edge which is labled maximum, in a way that each vertex has only one edge. Two subgraphes of "fig. 2" are shown in "fig. 3" and "fig. 4". To compute the weight of the graph ,we select the less degree of edges .

For example the weight of graph G1 in "fig. 3", is S and the weight of graph G2 in "fig. 4" is P. The result of matching two services is the graph which has the higher weight (in our example graph G2 that has the weight P). We do all these works for inputs, too.

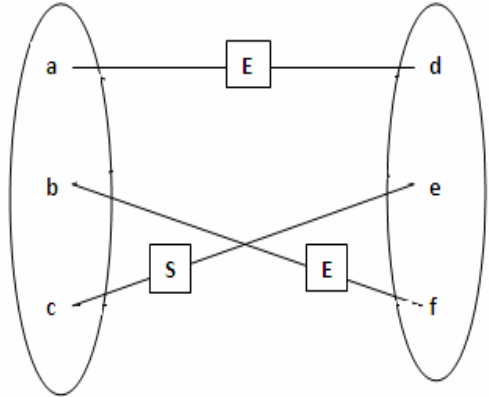

**Fig. 3.** Shows the matching subgraph G1

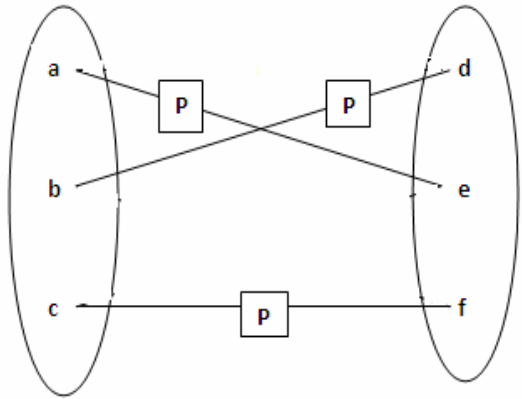

**Fig. 4.** Shows the matching subgraph G2

### 3.2  Our Matching Algorithm

If the result of matching two services` outputs is OUTSIM and the result of matching two services inputs is INSIM, the whole result of matching two services is "result" that obtains by the following algorithm . The input of our algorithm is C which is the crashed service. The outputs of our algorithm are "bestsrv" and "best result" which are the best service and the matching rate, respectively.

```
FIND MATCH (C , bestsrv, best result );
bestsrv = first service;
best result = F;

for all services in repository  (ADV)  do
    for all output parameters of  C do
         Case ( C_out , ADV_out ) ;
    make a bipartite graph for outputs ;
    OUTSIM = minimum edge in bipartite graph;

    for all input parameters of C do
         Case ( C_in , ADV_in );
    make a bipartite graph for inputs ;
    INSIM= minimum edge in bipartite graph ;

    reselt = E ;
    if   (OUTSIM=F or INSIM=F ) then
         reselt = F
              else if   (OUTSIM=S or INSIM=S ) then
                     reselt = S
                     else if   (OUTSIM=P or INSIM=P )
              then
                          reselt = P ;

if result > best result then
    best result = result ;
    bestsrv = ADV ;
if best result =E then quit ;
```

## 3.3  Complexity Analysis

In computing the complexity of semantic web service matching, a lot of factors have Interference such as number of services in the repository, number of input parameters and number of output parameters. In equal situation with other algorithms, we don`t consider the number of input and output parameters. So for computing time complexity of an algorithm, just the number of advertised services is important. Now if N is the number of advertised services in the repository, we choose the first service as the best one. After computing each advertised service`s similarity, If this one`s similarity rate is higher than the best one, then this service is chosen as the best and so on.

Whenever it finds the service by the similarity rate E, the work is finishing. Therefore the complexity of our algorithm is of O (N).

## 4   An Example

For better explanation of our algorithm, this section describes a simple scenario of matching web services.

If service C with these inputs and outputs crashes:
Inputs: (officer ID, company name)
Outputs: (name, address, phone number)

This service takes the name of a company and one of its officer`s ID. Now we want to find a service for substituting it. For example, one of the services in UDDI is ADV that has these inputs and outputs:

Inputs: (customer name, member ID)
Output: (name, mobile number, add)

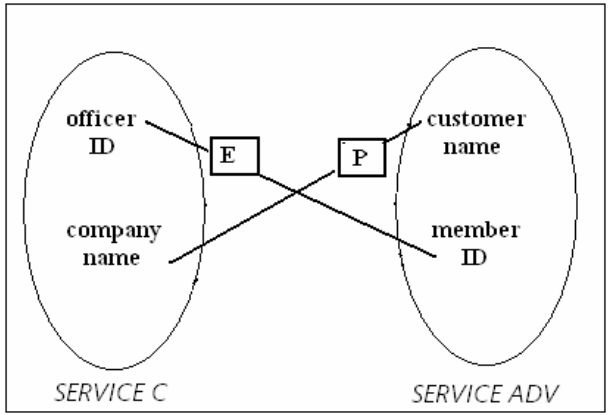

**Fig. 5.** Shows a bipartite graph for inputs

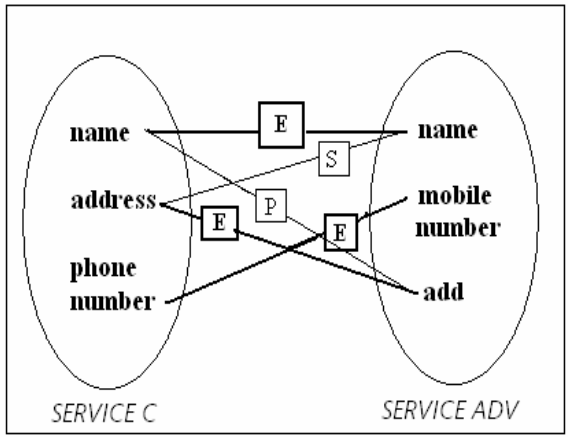

**Fig. 6.** Shows a bipartite graph for outputs

The bipartite graph for inputs is shown in "fig. 5". According to our ontology, the INSIM will be P.

The bipartite graph for outputs is shown in "fig. 6". According to our ontology, the OUTSIM will be E.

According to these two graphs, the result is P.

## 5   Discussion

For comparing two services, many languages are used such as : WSDL, OWL, OIL[3], DAML[4],  and etc. We used the service repository (UDDI) to do the semantic matching of web services. The complexity of our algorithm is at most O(N). It means in most cases with less than N computation we can get the similar service and among the other algorithms, it is one of the most precise algorithms. Always it considers the minimum similarity between two services. Therefore if it concludes the EXACT similarity, really has found the best one. Some of other works compute the average similarity or some others say two services are exact similar if one of the input or output has the result of exact. It is obvious that sometimes our algorithm may not find a match for the crashed web service, because we don`t sacrifice the accuracy. Some other algorithms have a lot of computation to find a numeric degree of matching. Although they compute the similarity exactly, but we think that our algorithm does the same work without need to do a lot of computation.

Some services have preconditions and effects. In this work we didn`t mention these two factors for simplicity, but in future work we surely consider them.

## 6   Conclusion and Future Work

In this paper we have identified the problem of semantic web matching and proposed a new algorithm to solve the problem. By the help of UDDI registry, which advertises services that all are described in OWL language, we could compare the functionality of a crashed service with others. We conclude that functionality matching gives better result than QOS or policy matching. We also proposed a new architecture for our work that shows the simplicity and accuracy of our work.

The bipartite graph played the main role of matching two semantic web services in our work. At the end We had computed the time complexity of our proposed algorithm ( O(N) ) which is minimum among other algorithms.

Some of web services have preconditions to do a work and then have some effects. By considering these two factors, we can obtain a better result, certainly.

Our future work is focused on improving the efficiency and accuracy of this algorithm by considering preconditions and effects of a service.

---

[3] Ontology Inference Layer.
[4] DARPA Agent Markup Language.

## References

1. Okutan, A.C., Kesim Cicekli, B.N.: A Monolithic Approach to Automated Composition of Semantic Web Services With The Event Calculus. knowledge-based systems 23, 440–454 (2010)
2. Abramowicz, W., Haniewicz, K., Kaczmarek, M., Zyskowski, D.: Architecture for Web Services Filtering and Clustering. In: 2nd IEEE international conference on internet and web applications and services (2007)
3. Segev, A.: Circular Context-Based Semantic Matching to Identify Web Service Composition. In: CSSSIA, Beijing, China (2008)
4. Yue, A.K., Liu, A.W., Wangb, X.: Zhou. A., Li, J.:Discovering Semantic Associations Among Web Services Based on the Qualitative Probabilistic Network. Expert systems with applications 36, 9082–9094 (2009)
5. Ivan, H., Akkiraju, R., Goodwin, R.: Learning Ontologies to Improve the Quality of Automatic Web Service Matching. In: IEEE international conference on web services (2007)
6. Doan, A., Madhavan, J., Domingos, P., Halevy, A.: Learning to Map Between Ontologies on the Semantic Web. Honolulu, Hawaii, USA (2002)
7. Ontology Matching Approaches in Semantic Web: a Survey. department of computing science, University of Alberta, Cdmonton, Canada
8. Nacer Talantikite, H., Aissani, D., Boudjlida, N.: Semantic Annotations for Web Services Discovery and Composition. Computer standards & interfaces 31, 1108–1117 (2009)
9. Fan, J.: Semantics-Based Web Service Matching Model. In: IEEE international conference on industrial informatics, pp. 323–328 (2006)
10. Verma, K., Akkiraju, R., Goodwin, R.: Semantic Matching of Web Service Policies
11. Chan Oh, S., Woon Yoo, J., Kil, H., Lee, D., Kumara, S.R.T.: Semantic Web-Service Discovery and Composition Using Flexible Parameter Matching. In: The 9th IEEE international conference on e-commerce technology and the 4th international conference on enterprise computing, e-commerce and e-services (2007)
12. Ben Mokhtar, S., Kaul, A., Georgantas, N., Issarny, V.: Towards Efficient Matching of Semantic Web Service Capabilities. In: International workshop on web services modeling and testing, pp. 137–152 (2006)
13. http://www.w3.org/
14. http://semanticweb.org/wiki/Main_Page
15. http://en.wikipedia.org/wiki/Bipartite_graph